\documentclass{article}

\PassOptionsToPackage{numbers, sort&compress}{natbib}
\usepackage[preprint]{neurips_2025}

\usepackage[utf8]{inputenc} % allow utf-8 input
\usepackage[T1]{fontenc}    % use 8-bit T1 fonts
\usepackage{hyperref}       % hyperlinks
\usepackage{url}            % simple URL typesetting
\usepackage{booktabs}       % professional-quality tables
\usepackage{graphicx}        % enhanced graphics support
\usepackage{amsfonts}       % blackboard math symbols
\usepackage{nicefrac}       % compact symbols for 1/2, etc.
\usepackage{microtype}      % microtypography
\usepackage{xcolor}         % colors
\usepackage{amsmath}
\usepackage{enumerate}
\usepackage{graphicx}
\usepackage{bm}         % 加粗符号
\usepackage{multirow, multicol}   % 多行合并
\usepackage{subcaption}
\usepackage{xcolor}
\usepackage[final]{changes}
\usepackage{float}
\usepackage{booktabs} % 三线表必备
\usepackage{multirow} % 跨行必备
\title{IndexTTS 2.5 Technical Report}

\author{%
  Index Speech Team \\
}

\begin{document}

\maketitle

\begin{abstract}
In \deleted{prior}\added{previous} work, we introduced IndexTTS 2, a zero-shot neural text-to-speech foundation model comprising two core components: a Transformer-based Text-to-Semantic (T2S) module and a non-autoregressive Semantic-to-Mel (S2M) module\deleted{,}\added{.} \deleted{which together enable faithful emotion replication and establish the first autoregressive duration-controllable generative paradigm.} \added{Jointly, the above two components enable faithful emotion replication and establish the first autoregressive duration-controllable generative paradigm.} Building upon this, we present IndexTTS 2.5, which significantly enhances multilingual coverage, inference speed, and overall synthesis quality through four key improvements: \textbf{1) Semantic Codec Compression:} \deleted{we reduce}\added{by reducing} the semantic codec frame rate from 50 Hz to 25 Hz, \deleted{halving sequence length and substantially lowering both training and inference costs}\added{the halved sequence length substantially decreases costs both training and inference progress}; \textbf{2) Architectural Upgrade:} we replace the U-DiT-based backbone of the S2M module with a more efficient Zipformer-based modeling architecture, achieving \deleted{notable parameter reduction}\added{less parameters} and faster mel-spectrogram generation \added{speed}; \textbf{3) Multilingual Extension:} we propose three explicit cross-lingual modeling strategies-boundary-aware alignment, token-level concatenation, and instruction-guided generation—establishing practical design principles for zero-shot multilingual emotional TTS that supports Chinese, English, Japanese, Spanish and Arabic, and enables robust emotion transfer even without target-language emotional training data; \textbf{4) Reinforcement Learning Optimization:} we apply Group Relative Policy Optimization (GRPO) in post-training of the T2S module, improving pronunciation accuracy and naturalness. Experiments show that IndexTTS 2.5 not only supports broader language coverage but also replicates emotional prosody in unseen languages \deleted{under}\added{with} the same zero-shot setting. IndexTTS 2.5 achieves a 2.28× improvement in real-time factor (RTF) while maintaining comparable word error rate (WER) and speaker similarity to IndexTTS 2. Audio demos are available at: https://index-tts.github.io/index-tts2-5.github.io/.

\end{abstract}

\section{Introduction}

In recent years, with the continuous advancement of generative frameworks, large-scale zero-shot text-to-speech (TTS) models have made remarkable progress, particularly in model architecture innovation \cite{valle, du2024cosyvoice, du2024cosyvoice2, chen2024f5, wang2024maskgct, deng2025indextts, wang2025spark}, audio quality enhancement \cite{anastassiou2024seed}, emotional expressiveness \cite{indextts2, voxcpm}, voice cloning \cite{seedvc, zhang2025vevo}, and multilingual coverage \cite{cosyvoice3, minimax-speech}. Currently, driven by trade-offs in framework capabilities and target application scenarios, mainstream large-scale TTS models have diverged into multiple technical paradigms, with their core differences primarily lying in model architecture design and the choice of intermediate representation modeling.

\begin{figure}[tp]
    \centering
    \includegraphics[width=1\textwidth]{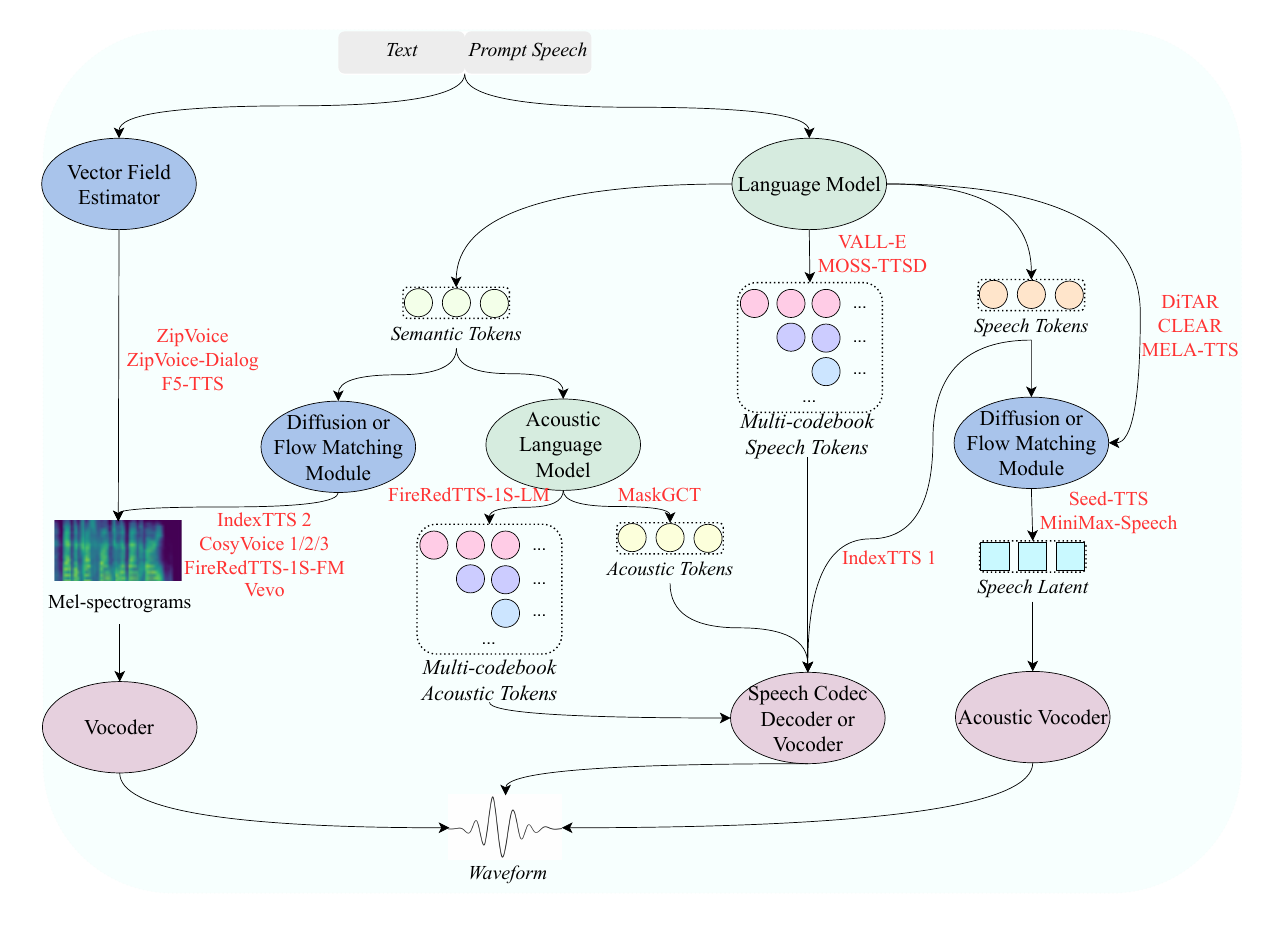}
    \caption{Schematic diagram of the technical approaches for current mainstream large-scale zero-shot speech synthesis models.}
    \label{fig:tree}
\end{figure}

As illustrated in Figure \ref{fig:tree}, modern large-scale zero-shot TTS architectures typically comprise a set of core components: a Transformer-based language model, a generative module based on diffusion or flow matching, a speech codec, and a neural vocoder. Concurrently, the choice of intermediate representation has evolved significantly—from early Mel-spectrograms to discrete speech tokens, and further diversified into semantic tokens, acoustic tokens, and continuous latent representations. Early large-scale TTS models, such as VALL-E \cite{valle}, pioneered the use of residual vector quantization (RVQ) \cite{audiolm, vq} to discretize speech into acoustic tokens and leveraged language models—inspired by the success of large language models—to generate these discrete tokens conditioned on text and a reference speech prompt \cite{mosstts}. This work demonstrated the effectiveness of leveraging large language models (LLMs) for zero-shot TTS, helping to establish a scalable generative framework that influenced subsequent research in neural speech synthesis. Building upon this foundation, several lines of model evolution have emerged. For instance, IndexTTS 1 \cite{deng2025indextts} discards the codec decoder entirely and instead trains a more powerful vocoder \cite{lee2022bigvgan} to reconstruct richer acoustic details directly from discrete tokens \cite{liao2024fish, casanova2024xtts}. In contrast, models like Seed-TTS \cite{anastassiou2024seed} and MiniMax-Speech \cite{minimax-speech} adopt a two-stage approach—first generating coarse speech tokens with a language model, then refining them via a diffusion- or flow matching–based module to learn fine-grained continuous latent representations, which are finally converted to waveforms by an acoustic vocoder. More recently, the speech token has been explicitly decomposed into semantic and acoustic tokens, the former encodes high-level linguistic content, prosody, and phonetic information, while the latter captures low-level acoustic characteristics. The prevailing architecture today employs a language model to generate semantic tokens, followed either by another language model that predicts acoustic tokens \cite{wang2024maskgct, guo2024fireredtts} or by a diffusion or flow matching module \cite{indextts2, du2024cosyvoice, du2024cosyvoice2, cosyvoice3, guo2024fireredtts} that directly generates Mel-spectrograms—a paradigm that has become the dominant approach in modern zero-shot TTS systems. Moreover, lightweight models \cite{e2tts} such as F5-TTS \cite{chen2024f5} and ZipVoice \cite{zipvoice, zipvoice-dialog} adopt a fully non-autoregressive approach to directly model the alignment between text and speech, enabling fast inference while maintaining high naturalness and audio quality. Recently, driven by deeper analyses of the trade-offs between discrete tokens and continuous representations—in terms of information fidelity, modeling efficiency, and generalization capability—an increasing number of works have shifted toward directly modeling continuous speech representations \cite{ditar, clear, mela-tts}. This paradigm aims to eliminate reliance on speech codec technologies and mitigate the inevitable information loss introduced by discretization.

To better disentangle emotion and speaker characteristics, IndexTTS 2 adopts a classic three-stage pipeline: it first employs a language model to generate discrete semantic tokens encoding emotion-related attributes such as pronunciation and prosody, then reconstructs speaker-specific acoustic details via a flow matching module and a neural vocoder. However, its limited multilingual coverage and slow inference hinder deployment in real-world, cross-lingual, or low-latency scenarios. To address these limitations, we present IndexTTS 2.5, a multilingual zero-shot TTS model that not only extends language coverage to Chinese, English, Japanese, Spanish and Arabic, but also preserves high-fidelity emotional expressiveness. Furthermore, through systematic architectural improvements and codec compression, IndexTTS 2.5 achieves substantial reductions in inference latency and computational cost, significantly enhancing its practical deployability. 

Our key practical enhancements are summarized as follows:
\begin{itemize}
\item[$\bullet$] To extend IndexTTS 2 to a broader set of languages, we first develop a multilingual data curation pipeline. Building on this pipeline, we propose three modeling strategies—boundary-aware alignment, token-level concatenation, and instruction-guided generation—for zero-shot cross-lingual TTS. Comprehensive evaluation not only confirms their individual effectiveness but also reveals complementary strengths among them, yielding practical design principles for future multilingual speech synthesis systems.
\item[$\bullet$] We propose a joint efficiency optimization—reducing the semantic codec frame rate from 50 Hz to 25 Hz and replacing U-DiT with the Zipformer architecture \cite{zipformer} in the flow-matching module—achieving a 2.28× reduction in RTF with no perceptible degradation in audio quality.
\item[$\bullet$] We apply GRPO \cite{grpo} to post-train the language model for semantic token generation, leading to consistent gains in prosodic naturalness and WER.
\end{itemize}

\section{The Multilingual Data Pipeline}
\begin{figure}[h]
    \centering
    \includegraphics[width=0.9\textwidth]{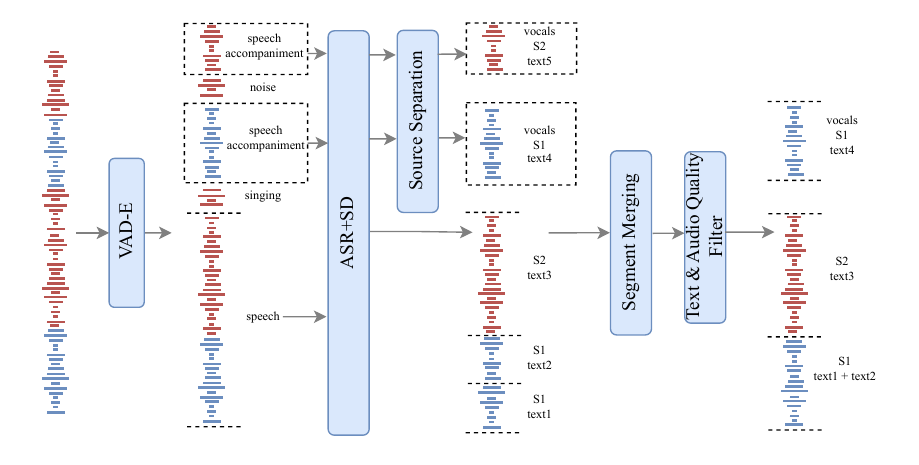}
    \caption{Multilingual data processing pipeline.}
    \label{fig:pipeline}
\end{figure}
To improve cross-lingual generalization in IndexTTS 2, we develop a multilingual data curation pipeline (Figure \ref{fig:pipeline}) that efficiently harvests high-quality speech–text pairs from publicly available sources such as online videos and audiobooks. The pipeline comprises several key components: voice activity detection (VAD) and segmentation, automatic speech recognition (ASR), speaker diarization, audio source separation, segment merging, and sample quality filtering.

\textbf{Voice activity detection and segmentation. } A voice activity detection model enhanced with event classification capability. For each audio segment, it estimates the proportion of speech, singing, accompaniment, and background noise.

\textbf{Automatic speech recognition. } An integrated ASR system that jointly performs speech recognition, speaker diarization, and punctuation restoration. The output for each short segment includes: transcribed text, speaker ID, and binary flags indicating the presence of multiple speakers, singing, or musical accompaniment.

\textbf{Source Separation. } Audio segments flagged as containing accompaniment are processed using the open-source Demucs toolkit \cite{demucs} to extract vocal tracks. To avoid potential audio degradation, source separation is applied selectively—only when necessary.

\textbf{Segment Merging. } Short ASR segments are merged into longer utterances based on speaker consistency, textual coherence, and inter-segment silence. Each resulting segment contains a single speaker and is constrained to a maximum duration of 25 seconds.

\textbf{Sample quality filtering. } A two-stage filtering module is applied to ensure data quality: audio quality filtering employs a pre-trained audio quality assessment model, while text quality filtering leverages transcripts from an ASR system to detect and discard low-fidelity or erroneous transcriptions.

\section{IndexTTS 2.5}

IndexTTS 2.5 retains the overall generation pipeline of IndexTTS 2, with targeted enhancements in four key components: multilingual text-to-semantic modeling, semantic codec frame rate compression, efficient semantic-to-mel conversion based on an improved backbone, and a reinforcement learning (RL)–based post-training mechanism for synthesis quality refinement. We describe the design and implementation of each module in the following subsections. 

\subsection{Multilingual Text-to-Semantic Module}

The text-to-semantic module in IndexTTS 2.5 retains the architecture of IndexTTS 2. Specifically, it takes a structured input in the form of $[c, p,  e_{\langle BT\rangle}, E_{text},  e_{\langle BA\rangle}, E_{sem}]$, let $c$ denote a global conditioning vector that captures speaker identity or emotional attributes, and $p$ an auxiliary embedding for fine-grained duration control. The text is represented as a sequence of embeddings $E_{text}$, while the target semantic tokens—derived by encoding the ground-truth speech with a semantic codec—are embedded as $E_{sem}$. To facilitate cross-modal alignment, special prefix tokens $e_{\langle BT\rangle}$ and $e_{\langle BA\rangle}$ are prepended to the text and semantic sequences, respectively.

\begin{figure}[h]
    \centering
    \includegraphics[width=0.9\textwidth]{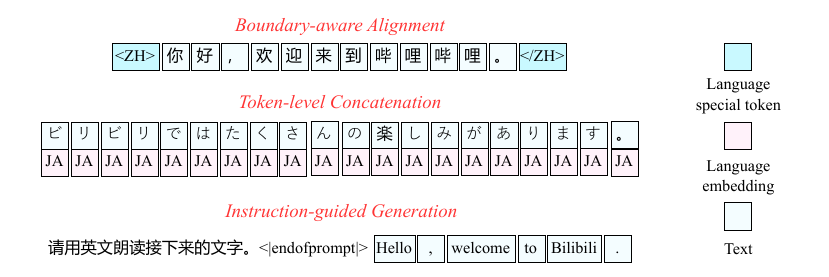}
    \caption{Illustration of three multilingual modelling strategies: boundary-aware alignment, token-level concatenation, and instruction-guided generation.}
    \label{fig:lang}
\end{figure}

In multilingual settings, character overlap across languages—such as the extensive use of Chinese characters in Japanese—can lead to cross-lingual confusion during zero-shot speech synthesis. As shown in Figure \ref{fig:lang}, to address this challenge, we propose three targeted strategies: boundary-aware alignment, token-level concatenation, and instruction-guided generation.

\subsubsection{Boundary-Aware Alignment Strategy}

Boundary-aware alignment is designed to enable the T2S module to learn correct pronunciation in multilingual settings, mitigating confusion caused by shared characters across languages. Specifically, we insert language-specific boundary tokens—such as <ZH> and </ZH> for Mandarin—around each input utterance to explicitly mark the start and end of a language segment. The input sequence here is constructed as $[c, p,  e_{\langle BT\rangle}, e_{\langle LID \rangle}, E_{text}, e_{\langle /LID \rangle}, e_{\langle BA\rangle}, E_{sem}]$. This is one of the simplest yet effective multilingual modeling strategies, providing clear signals for language boundaries during training and inference. However, due to the hallucination tendency inherent in autoregressive generation, its control over pronunciation degrades on longer utterances, occasionally leading to mispronunciations.

\subsubsection{Token-Level Concatenation Strategy}

We propose a token-level strict concatenation strategy, in which each input text token embedding is fused with a language-specific categorical embedding. This provides explicit per-token linguistic supervision, enabling the model to disambiguate pronunciation based on the token’s linguistic origin—particularly critical for characters shared across languages. As a result, cross-lingual homograph-induced mispronunciations are effectively suppressed. However, practical deployment necessitates a front-end module capable of fine-grained, token-level language identification, rendering system integration more complex than with the boundary-aware alignment approach.

\subsubsection{Instruction-Guided Generation Strategy}

To better exploit the text comprehension capabilities of the underlying language model, we adopt an instruction-guided generation strategy: structured natural-language instructions are prepended to the input text to explicitly condition the model on the target language. To enhance training diversity and promote contextual awareness, we design a set of prompt templates (e.g., as shown in Figure \ref{fig:lang}) and randomly sample one as the prefix during training. The resulting input sequence follows the format $[c, p,  e_{\langle BT\rangle}, e_{instr}, e_{\langle EOP \rangle} , E_{text}, e_{\langle BA\rangle}, E_{sem}]$, where $e_{instr}$ denotes the instruction embedding and EOP marks the end of the prompt. This design eliminates the need for external language tagging modules at inference time, enabling fully self-contained multilingual generation. However, in code-switching scenarios with mixed-language utterances, portions of the instruction may provide limited utility for speech synthesis, introducing mild redundancy in the input sequence.

\subsection{Efficiency-Oriented System Design}
To reduce computational overhead and latency in speech generation, we incorporate two complementary optimizations across the pipeline: 1) semantic codec compression, which reduces the frame rate from 50 Hz to 25 Hz, halving the length of the target semantic token sequence; and 2) architectural streamlining, which replaces the U-DiT backbone in the S2M module with a more parameter-efficient Zipformer architecture \cite{zipformer}. The shortened token sequence lowers memory consumption and computational load during both training and inference, while the compact Zipformer architecture further reduces model complexity. These design choices collectively improve inference efficiency with minimal impact on synthesis quality.

\subsection{Reinforcement Learning for Post-Training}

We fine-tune the text-to-semantic model using GRPO with a preference-based objective. For each input text prompt—particularly in multilingual or code-switching scenarios—we generate four diverse speech candidates via stochastic decoding. These candidates are ranked by a reward signal that approximates intelligibility through the word error rate produced by a frozen ASR system, and speaker similarity through a speaker verification model. Under the GRPO framework, the policy is updated to increase the relative likelihood of higher-reward (i.e., lower WER and higher SS) generations compared to lower-reward ones.

\subsection{G2P Strategy} 

The original G2P framework was extended to accommodate Chinese, English, and Japanese. For Chinese and English, we restricted the lexicon to entries with unambiguous pronunciations, utilizing the CMU Pronouncing Dictionary for the latter. n the case of Japanese, Kanji characters were randomly sampled and converted to Katakana via the fugashi \cite{mccann-2020-fugashi}. Since Japanese readings depend on lexical context, direct replacement may occasionally introduce unnatural pronunciations. This issue can be alleviated by performing the conversion within segmented text units rather than across the entire sentence.

\section{Experiments and Evaluation}
\label{others}

\subsection{Training and Evaluation Datasets}

We train a multilingual text-to-speech model covering Mandarin, English, Japanese, and Spanish, using a total of approximately 100K hours of speech data. The corpus comprises 30K hours of Mandarin, 25K hours of English, 42K hours of Japanese, 22.9K hours of Spanish and 20k hours of Arabic.

Mandarin and English data are derived primarily from the Emilia dataset \cite{he2024emilia}, augmented with audiobooks and commercially licensed recordings. For Japanese, 1.7K hours originate from Emilia \cite{he2024emilia}, with the remainder obtained through commercial licensing. Spanish data is aggregated from multiple public sources: Common Voice \cite{commonvoice}, the Argentinian Spanish Speech Dataset \cite{argentinian}, the TEDx Spanish Corpus \cite{tedx}, the Crowdsourced High-Quality Spanish Speech Dataset \cite{crowd}, Multilingual LibriSpeech \cite{libri}, and LEMAS \cite{zhao2026lemas}. For Arabic, all were obtained through commercial licensing. To support expressive synthesis, we further incorporate 135 hours of emotional speech, including 29 hours from the Emotional Speech Dataset (ESD) \cite{esddataset} and 106 hours of high-quality emotional recordings acquired under commercial license.

% For evaluation on Mandarin and English, we adopt the SeedTTS test-zh and SeedTTS test-en benchmarks \cite{anastassiou2024seed}: test-zh comprises 2,000 utterances from DiDiSpeech \cite{guo2021didispeech}, and test-en contains 1,000 utterances from Common Voice \cite{commonvoice}. We construct language-specific test sets for Japanese and Spanish, denoted IndexTTS test-ja and IndexTTS test-es, each combining 500 and 300 utterances, respectively, from Common Voice \cite{commonvoice} and FLEURS \cite{fluers}. To evaluate emotion expressiveness in these two languages, we further introduce IndexTTS test-ja-emo and IndexTTS test-es-emo, curated entirely from publicly available online speech resources.

% For Mandarin and English, we use the Seed-TTS-Eval \cite{anastassiou2024seed} and CV3-Eval \cite{cosyvoice3} benchmarks. For Japanese and Spanish, we use the corresponding CV3-Eval evaluation sets. To further evaluate emotion cloning, we use the emotion-zeroshot benchmark from CV3-Eval and additionally introduce three in-house benchmarks: IndexTTS test-ja-emo, IndexTTS test-es-emo, and IndexTTS test-ar-emo. All three datasets are curated from publicly available speech resources.

To evaluate the multilingual modeling approaches, we conducted ablation studies using the Seed-TTS-Eval benchmark for Chinese and English, alongside in-house test sets for Japanese and Spanish. Zero-shot and cross-lingual performances are evaluated on the multilingual CV3-Eval benchmark. For Arabic, where no standard benchmark is available, we construct proprietary zero-shot and cross-lingual sets. We also built our own Spanish cross-lingual set.

\subsection{Evaluation Metrics and Baselines} 

We perform a comprehensive evaluation on four dimensions: word error rate (WER), speaker similarity (SS), emotional similarity (ES), and mean opinion score (MOS). WER is computed using language-specific ASR models: FunASR \cite{funasr} for Chinese, and Whisper-large-v3 \cite{whisper} for the others. Speaker similarity is measured by extracting speaker embeddings from the reference prompt and the generated utterance using FunASR’s pretrained speaker verification model \cite{funasr}, then computing their cosine similarity. Emotional similarity is similarly evaluated via cosine similarity between emotion embeddings extracted with emotion2vec \cite{emotion2vec}. MOS is obtained through subjective listening tests, where native speakers rate naturalness on a 1–5 scale (1: poor, 5: excellent).

We compare our model against state-of-the-art zero-shot multilingual TTS systems, including CosyVoice3 \cite{cosyvoice3}, FireRedTTS-2 \cite{guo2024fireredtts}, Fish Audio S2 Pro \cite{liao2026fishaudios2technical}, Qwen3-TTS \cite{Qwen3-TTS}, VoxCPM2 \cite{zhou2026voxcpm2}, Moss-TTS \cite{gong2026mossttstechnicalreport}, and the original IndexTTS 2 \cite{indextts2}.

% We conduct a comprehensive evaluation across four dimensions: WER, speaker similarity (SS), emotional similarity (ES), and mean opinion score (MOS). WER is computed using language-specific ASR models: FunASR \cite{funasr} for SeedTTS test-zh, and Whisper-large-v3 \cite{whisper} for SeedTTS test-en, IndexTTS test-ja, and IndexTTS test-es. Speaker similarity is measured by extracting speaker embeddings from the reference prompt and the generated utterance using FunASR’s pretrained speaker verification model \cite{funasr}, then computing their cosine similarity. Emotional similarity is similarly evaluated via cosine similarity between emotion embeddings extracted with emotion2vec \cite{emotion2vec}. MOS is obtained through subjective listening tests, where native speakers rate naturalness on a 1–5 scale (1: poor, 5: excellent).

% We compare our model against state-of-the-art zero-shot multilingual TTS systems, including CosyVoice3 \cite{cosyvoice3}, FireRedTTS-2 \cite{guo2024fireredtts}, and the original IndexTTS 2 \cite{indextts2}.

\subsection{Comparative Analysis of Multilingual Modelling Methods}
\begin{table}[t]
\centering
\caption{Performance comparison of various multilingual modeling methods (Boundary-Aware Alignment, Token-Level Concatenation, Instruction-Guided Generation) on Mandarin, English, Japanese, Spanish test sets.}
%Specifically, WER values are reported as percentages.
\label{tab:multilingual-methods} % It's good practice to add a label
\begin{tabular}{@{}lcccc@{}}
\toprule
\textbf{Dataset} & \textbf{Model} & \textbf{SS$\uparrow$} & \textbf{WER(\%)$\downarrow$}  \\
\midrule
% \multirow{7}{*}{\textbf{LibriSpeech test-clean}}
\multirow{3}{*}{\begin{tabular}{@{}c@{}}\textbf{SeedTTS} \\ \textbf{test-zh}\end{tabular}}
& Boundary-Aware Alignment & 0.797 & 1.602  \\
& Token-Level Concatenation & \textbf{0.804} & \textbf{1.119}  \\
& Instruction-Guided Generation  & 0.798 & 1.613 \\
\midrule % Separator between dataset blocks
% SeedTTS test-en block
% \multirow{7}{*}{\textbf{SeedTTS test-en}}
\multirow{3}{*}{\begin{tabular}{@{}c@{}}\textbf{SeedTTS} \\ \textbf{test-en}\end{tabular}}
& Boundary-Aware Alignment & 0.820 & 3.577 \\
& Token-Level Concatenation & \textbf{0.823} & \textbf{3.253} \\
& Instruction-Guided Generation  & 0.819 & 3.314 \\
\midrule % Separator
% SeedTTS test-zh block
% \multirow{7}{*}{\textbf{SeedTTS test-zh}}
\multirow{3}{*}{\begin{tabular}{@{}c@{}}\textbf{IndexTTS} \\ \textbf{test-ja}\end{tabular}}
& Boundary-Aware Alignment & 0.672 & 10.444  \\
& Token-Level Concatenation & \textbf{0.745} & 10.498  \\
& Instruction-Guided Generation & 0.667 & \textbf{9.226}  \\
\midrule % Separator
% AIShell-1 test block
% \multirow{7}{*}{\textbf{AIShell-1 test}}
\multirow{3}{*}{\begin{tabular}{@{}c@{}}\textbf{IndexTTS} \\ \textbf{test-es}\end{tabular}}
& Boundary-Aware Alignment & 0.784 & 5.832 \\
& Token-Level Concatenation & \textbf{0.788} & \textbf{5.200}  \\
& Instruction-Guided Generation & 0.779 & 5.399 \\
\bottomrule
\end{tabular}
\end{table}
To ensure a fair comparison, all three multilingual modeling strategies—Boundary-Aware Alignment, Token-Level Concatenation, and Instruction-Guided Generation—are trained under identical conditions: using the same multilingual dataset, batch size, optimizer, learning rate schedule, and hardware resources, with training terminated after 100K steps.

We then conduct a comprehensive evaluation of these models across four test sets covering Mandarin, English, Japanese, and Spanish. As shown in Table \ref{tab:multilingual-methods}, Token-Level Concatenation consistently achieves the highest speaker similarity across all languages, indicating its strong capability in preserving speaker identity during cross-lingual synthesis. This is attributed to the explicit per-token language conditioning, which enables fine-grained alignment between input text and speaker characteristics. Token-Level Concatenation achieves the lowest WER among all evaluated strategies on Mandarin, English, and Spanish, with 1.119\% on Seed-TTS-Eval test-zh, 3.253\% on Seed-TTS-Eval test-en, and 5.200\% on IndexTTS test-es. However, on the IndexTTS test-ja, its WER is slightly higher than that of Instruction-Guided Generation, suggesting that the rigid concatenation may introduce subtle misalignments in morphologically complex languages. In contrast, Instruction-Guided Generation achieves the lowest WER on Japanese, likely due to its ability to leverage contextual cues from structured prompts for better phonetic disambiguation.

Notably, while Boundary-Aware Alignment performs competitively on Mandarin and English, it underperforms on Japanese and Spanish, particularly in WER, highlighting its limitation in handling languages with rich morphology or non-Latin scripts. The consistent improvement of Token-Level Concatenation across both SS and WER underscores its robustness and effectiveness in zero-shot multilingual TTS. While Instruction-Guided Generation offers greater flexibility at inference time, its performance is sensitive to prompt design and requires careful engineering, whereas Token-Level Concatenation provides more stable and reliable results with minimal additional overhead.

\subsection{Voice Cloning Evaluation}
\begin{table}[H]
\centering
\caption{Zero-shot TTS evaluation results. WER (\%) $\downarrow$ and Speaker Similarity (SS) $\uparrow$ are reported. $^\dagger$Results cited from the original paper.}
\label{tab:zeroshot}
% ============================
% Table 1: Zero-shot TTS
% ============================
\resizebox{\textwidth}{!}{
\begin{tabular}{lc cc cc cc cc cc cc}
\toprule
\multirow{2}{*}{\textbf{Model}} & \multirow{2}{*}{\textbf{Params}}
  & \multicolumn{2}{c}{\textbf{test-zh}}
  & \multicolumn{2}{c}{\textbf{test-en}}
  & \multicolumn{2}{c}{\textbf{test-es}}
  & \multicolumn{2}{c}{\textbf{test-ja}}
  & \multicolumn{2}{c}{\textbf{test-ar}}
  & \multicolumn{2}{c}{\textbf{Average}} \\
\cmidrule(lr){3-4}\cmidrule(lr){5-6}\cmidrule(lr){7-8}\cmidrule(lr){9-10}\cmidrule(lr){11-12}\cmidrule(lr){13-14}
& & WER & SS & WER & SS & WER & SS & WER & SS & WER & SS & WER & SS \\
\midrule
VoxCPM2          & 2B   & 3.88  & 74.99 & 5.13  & \textbf{}{71.57} & 5.49  & 74.67 & 6.69  & 72.90 & 14.94 & 65.99 & 7.22  & 72.02 \\
OmniVoice        & 0.8B & 3.41  & 72.99 & \textbf{3.62}  & 70.13 & 3.52  & 74.14 & 5.38  & 70.49 & 17.88 & 64.22 & 6.76  & 70.39 \\
Moss-TTS 1.5     & 8B   & 4.02  & 72.68 & 4.45  & 67.46 & 3.83  & 71.75 & 10.97 & 68.71 & 23.71 & 62.21 & 9.40  & 68.56 \\
CosyVoice3-0.5B  & 0.5B & 3.84 & \textbf{80.01} & 4.88 & \textbf{74.16} & 4.04 & \textbf{78.85} & -- & \textbf{76.36} & -- & -- \\
CosyVoice3-1.5B  & 1.5B & 3.91$^\dagger$ & --    & 4.99$^\dagger$ & --    & 4.47$^\dagger$ & --    & 7.57$^\dagger$ & --    & --    & --    & --    & --    \\
% CosyVoice3-0.5B  & 0.5B & 3.84  & \textbf{80.01} & 4.88  & \textbf{74.16} & 4.04  & \textbf{78.85} & 57.78 & \textbf{76.36} & --    & --    & --    & --    \\
FireRedTTS-2     & 1.5B & 8.22  & 68.10 & 14.92 & 56.93 & --    & --    & --    & --    & --    & --    & --    & --    \\
Fish Audio S2 Pro & 4B  & 3.62  & 67.79 & 3.83  & 61.66 & 2.93  & 67.44 & \textbf{5.15}  & 66.15 & 14.15 & 59.43 & \textbf{5.94}  & 64.49 \\
Qwen3-TTS        & 1.7B & \textbf{3.27}  & 73.02 & 5.06  & 67.17 & \textbf{2.87}  & 73.17 & 5.89  & 70.18 & --    & --    & --    & --    \\
\midrule
    IndexTTS2.5      & 0.8B & 4.36  & 77.10 & 5.12  & 68.06 & 3.75  & 76.39 & 5.66  & 74.62 & 14.88 & 69.74 & 6.75  & 73.18 \\
IndexTTS2.5-RL   & 0.8B & 3.93  & 77.92 & 3.89  & 67.79 & 3.33  & 76.68 & 5.30  & 75.41 & \textbf{13.58} & \textbf{70.36} & 6.00  & \textbf{73.63} \\
\bottomrule
\end{tabular}}

\centering
\caption{Cross-lingual TTS evaluation (Chinese prompt $\rightarrow$ target language). WER (\%) $\downarrow$ and Speaker Similarity (SS) $\uparrow$ are reported. $^\dagger$Results cited from the original paper.}
\label{tab:crosslingual}
\resizebox{\textwidth}{!}{
\begin{tabular}{lc cc cc cc cc cc}
\toprule
\multirow{2}{*}{\textbf{Model}} & \multirow{2}{*}{\textbf{Params}}
  & \multicolumn{2}{c}{\textbf{zh$\rightarrow$en}}
  & \multicolumn{2}{c}{\textbf{zh$\rightarrow$es}}
  & \multicolumn{2}{c}{\textbf{zh$\rightarrow$ja}}
  & \multicolumn{2}{c}{\textbf{zh$\rightarrow$ar}}
  & \multicolumn{2}{c}{\textbf{Average}} \\
\cmidrule(lr){3-4}\cmidrule(lr){5-6}\cmidrule(lr){7-8}\cmidrule(lr){9-10}\cmidrule(lr){11-12}
& & WER & SS & WER & SS & WER & SS & WER & SS & WER & SS \\
\midrule
VoxCPM2          & 2B   & 4.48  & 64.25 & 16.38 & 64.89 & 11.84 & 71.54 & 11.09 & 67.62 & 10.95 & 67.08 \\
OmniVoice        & 0.8B & 3.74  & 64.91 & 5.84  & 62.08 & 9.09  & 69.06 & 19.80 & 65.27 & 9.62  & 65.33 \\
Moss-TTS 1.5     & 8B   & 6.13  & 59.23 & \textbf{4.32}  & 56.63 & 11.52 & 65.54 & 17.03 & 62.93 & 9.75  & 61.08 \\
CosyVoice3-0.5B  & 0.5B & \textbf{3.23}  & 62.79    & 4.58    & 64.04    & -- & --    & --    & --    & --    & --    \\
CosyVoice3-1.5B  & 1.5B & 4.32  & --    & --    & --    & 13.70$^\dagger$ & --    & --    & --    & --    & --    \\
% CosyVoice3-0.5B  & 0.5B & \textbf{3.23}  & 62.79 & 4.59  & 64.04 & 53.65 & 72.46 & --    & --    & --    & --    \\
FireRedTTS-2     & 1.5B & 9.34  & 53.19 & 12.25 & 58.31 & 19.05 & 64.12 & --    & --    & --    & --    \\
Fish Audio S2 Pro & 4B  & 4.14  & 55.89 & 4.46  & 55.57 & 10.48 & 61.74 & 14.49 & 59.80 & 8.39  & 58.25 \\
Qwen3-TTS        & 1.7B & 5.74  & 63.04 & 5.15  & \textbf{68.02} & 36.09 & 65.71 & --    & --    & --    & --    \\
\midrule
IndexTTS2.5      & 0.8B & 3.62  & 63.83 & 5.17  & 65.48 & 6.57  & 74.16 & \textbf{9.51} & 71.02 & 6.22  & 68.62 \\
IndexTTS2.5-RL   & 0.8B & 3.55  & \textbf{67.47} & 4.86  & 64.47 & \textbf{6.38}  & \textbf{75.82} & 9.89  & \textbf{73.05} & \textbf{6.17}  & \textbf{70.20} \\
\bottomrule
\end{tabular}}

\end{table}
We first evaluate zero-shot TTS performance across five target languages. As shown in Table \ref{tab:zeroshot}, despite having only 0.8B parameters, IndexTTS 2.5 achieves the strongest overall performance among all compared models, with the lowest average WER (6.75) and the best average speaker similarity (73.18). While CosyVoice3-0.5B achieves higher speaker similarity on several individual language benchmarks and Fish Audio S2 Pro achieves lower average WER, IndexTTS 2.5 provides a more balanced performance, maintaining competitive speaker similarity while achieving consistently low WER across all evaluated languages. After RL fine-tuning, IndexTTS 2.5-RL further reduces the average WER from 6.75 to 6.00 and improves the average speaker similarity from 73.18 to 73.63, with consistent gains observed on English, Japanese, and Arabic.

We further evaluate cross-lingual TTS generation using Chinese speech prompts and target texts in English, Spanish, Japanese, and Arabic. Table \ref{tab:crosslingual} presents the corresponding results. IndexTTS 2.5 achieves the strongest overall cross-lingual performance, obtaining the lowest average WER (6.22) and the best average speaker similarity (68.62). Although CosyVoice3-0.5B achieves the lowest WER on the Chinese-to-English benchmark, IndexTTS 2.5 shows more consistent performance across different target languages, particularly on Japanese and Arabic. RL fine-tuning further improves the average WER to 6.17 and increases the average speaker similarity to 70.20, demonstrating additional gains in cross-lingual voice cloning.

\subsection{Multilingual Emotional TTS Evaluation}
We evaluate emotion cloning \added{performance} using \deleted{both} our in-house emotional speech benchmark and the CV3-Eval emotion-zeroshot benchmark\added{, comparing IndexTTS 2.5 with CosyVoice 3.} \added{As shown in Table \ref{tab:emotion_comparison}, on the internal Japanese and Spanish emotional test sets, IndexTTS 2.5 consistently produces higher emotion similarity (ES) while maintaining competitive WER and SS. It also receives a higher MOS score than CosyVoice 3. For the Arabic evaluation, Japanese speeches are used as the prompt audio due to the lack of matched prompt data. Although this cross-lingual setting could result in a lower speaker similarity score, the model still achieves a MOS of 4.18, demonstrating that it can effectively transfer emotional characteristics across languages.} \deleted{IndexTTS 2.5 is compared against CosyVoice 3 using both subjective and objective metrics. }\deleted{It is worth noting that emotion training is performed using only Chinese and English emotional speech data. Nevertheless, IndexTTS 2.5 generalizes effectively to all supported languages, demonstrating robust emotion transfer and emotion–speaker disentanglement capabilities beyond the languages seen during training.} \added{Note that for Arabic, Japanese speech was used as the prompt, potentially leading to a lower SS.} \added{In particular, the model achieves strong zero-shot emotion transfer on unseen languages, even though it was trained using only Chinese and English emotional speech data.}

\deleted{As shown in Table \ref{tab:emotion_comparison}, on the internal Japanese and Spanish emotional test sets, IndexTTS 2.5 consistently produces higher emotion similarity (ES) while maintaining competitive WER and SS. It also receives a higher MOS score than CosyVoice 3. For the Arabic evaluation, Japanese speeches are used as the prompt audio due to the lack of matched prompt data. Although this cross-lingual setting could result in a lower speaker similarity score, the model still achieves a MOS of 4.18, demonstrating that it can effectively transfer emotional characteristics across languages.}

\begin{table}[H] % 1. 用 H 强行锁死在当前位置
\centering
\footnotesize % 2. 将内部字号微调为 footnotesize（比 small 稍微再小一点点，非常适合多列表格）
\setlength{\tabcolsep}{2.8pt} % 3. 极限压缩列与列之间的间距（默认是 6pt 左右，这里压到 2.8pt），让表格横向自然变窄
\setlength{\abovecaptionskip}{4pt} % 4. 紧凑标题上方距离
\setlength{\belowcaptionskip}{2pt} % 5. 紧凑标题下方距离
\caption{\deleted{MOS scores and objective}\added{Objective and subjective} metrics on Japanese, Spanish, and Arabic emotional test sets.\deleted{ Note that for Arabic, we used Japanese as prompt wave, which may indicate a lower SS.}}
\label{tab:emotion_comparison}
\begin{tabular}{l c c c c c c c c c c c c}
\toprule
\multirow{2}{*}{\textbf{Model}} & \multicolumn{4}{c}{\textbf{IndexTTS test-es-emo}} & \multicolumn{4}{c}{\textbf{IndexTTS test-ja-emo}} & \multicolumn{4}{c}{\textbf{IndexTTS test-ar-emo}} \\
\cmidrule(lr){2-5} \cmidrule(lr){6-9} \cmidrule(lr){10-13}
& \textbf{SS$\uparrow$} & \textbf{WER$\downarrow$} & \textbf{ES$\uparrow$} & \textbf{MOS$\uparrow$} 
& \textbf{SS$\uparrow$} & \textbf{WER$\downarrow$} & \textbf{ES$\uparrow$} & \textbf{MOS$\uparrow$} 
& \textbf{SS$\uparrow$} & \textbf{WER$\downarrow$} & \textbf{ES$\uparrow$} & \textbf{MOS$\uparrow$} \\
\midrule
CosyVoice 3 & \textbf{0.811} & \textbf{8.574} & 0.831 & 3.96 & \textbf{0.765} & 10.28 & 0.844 & 3.52 & - & - & - & - \\
IndexTTS2.5 & 0.796 & 9.035 & \textbf{0.922} & \textbf{4.15} & 0.759 & \textbf{7.387} & \textbf{0.896} & \textbf{3.93} & 0.698 & 2.91 & 0.869 & 4.18 \\
\bottomrule
\end{tabular}
\end{table}

% \begin{table}[H]
% \centering
% \footnotesize
% \setlength{\tabcolsep}{4pt}

% \caption{Emotion perception evaluation on the CV3-Eval Emotion-ZeroShot benchmark.}
% \label{tab:emotion_perception}

% \begin{tabular}{llcc}
% \toprule
% \multicolumn{2}{c}{\textbf{Evaluation}} & 
% \multicolumn{2}{c}{\textbf{Model}} \\
% \cmidrule(lr){1-2} \cmidrule(lr){3-4}
% \textbf{Category} & \textbf{Metric} &
% \textbf{CosyVoice 3} & \textbf{IndexTTS 2.5} \\
% \midrule

% \multirow{4}{*}{Overall}
% & Accuracy (zh) & 0.467 & \textbf{0.713} \\
% & Accuracy (en) & 0.567 & \textbf{0.673} \\
% & Weighted-F1 (zh) & 0.585 & \textbf{0.790} \\
% & Weighted-F1 (en) & 0.670 & \textbf{0.759} \\

% \midrule

% \multirow{6}{*}{Emotion F1}
% & Angry (zh) & 0.457 & \textbf{0.705} \\
% & Angry (en) & 0.705 & \textbf{0.800} \\
% & Happy (zh) & 0.688 & \textbf{0.931} \\
% & Happy (en) & 0.667 & \textbf{0.776} \\
% & Sad (zh) & 0.611 & \textbf{0.734} \\
% & Sad (en) & 0.640 & \textbf{0.701} \\

% \midrule

% \multirow{4}{*}{Intensity Accuracy}
% & High (zh) & 0.720 & \textbf{0.813} \\
% & High (en) & 0.720 & \textbf{0.813} \\
% & Low (zh) & 0.213 & \textbf{0.613} \\
% & Low (en) & 0.413 & \textbf{0.533} \\

% \bottomrule
% \end{tabular}
% \end{table}

We further evaluate emotion perception using the CV3-Eval Emotion-ZeroShot benchmark\deleted{. As} \added
{, as }summarized in Tables \deleted{\ref{tab:emotion_perception}}\added{\ref{tab:emotion_perception_lang_first}}, \deleted{reports the overall emotion perception performance on the CV3-Eval Emotion-ZeroShot benchmark.} IndexTTS 2.5 consistently outperforms CosyVoice 3 on both Chinese and English across all reported metrics. The overall accuracy improves from 0.467 to 0.713 in Chinese and from 0.567 to 0.673 in English, while the weighted F1-score increases from 0.585 to 0.790 and from 0.670 to 0.759, respectively. The \deleted{following}\added{above} analyses further break down these improvements by emotion category and intensity level.

\begin{table}[H]
\centering
\footnotesize
\setlength{\tabcolsep}{9.5pt}
\caption{Emotion perception evaluation on the CV3-Eval Emotion-ZeroShot benchmark.}
\label{tab:emotion_perception_lang_first}

\begin{tabular}{ll cc ccc cc}
\toprule
\multirow{2}{*}{\textbf{Lang}} & \multirow{2}{*}{\textbf{Model}} & \multicolumn{2}{c}{\textbf{Overall}} & \multicolumn{3}{c}{\textbf{Emotion F1}} & \multicolumn{2}{c}{\textbf{Intensity Acc}} \\
\cmidrule(lr){3-4} \cmidrule(lr){5-7} \cmidrule(lr){8-9}
& & \textbf{Acc} & \textbf{W-F1} & \textbf{Angry} & \textbf{Happy} & \textbf{Sad} & \textbf{High} & \textbf{Low} \\
\midrule

\multirow{2}{*}{\textbf{zh}} 
& CosyVoice 3  & 0.467 & 0.585 & 0.457 & 0.688 & 0.611 & 0.720 & 0.213 \\
& IndexTTS 2.5 & \textbf{0.713} & \textbf{0.790} & \textbf{0.705} & \textbf{0.931} & \textbf{0.734} & \textbf{0.813} & \textbf{0.613} \\

\midrule

\multirow{2}{*}{\textbf{en}} 
& CosyVoice 3  & 0.567 & 0.670 & 0.705 & 0.667 & 0.640 & 0.720 & 0.413 \\
& IndexTTS 2.5 & \textbf{0.673} & \textbf{0.759} & \textbf{0.800} & \textbf{0.776} & \textbf{0.701} & \textbf{0.813} & \textbf{0.533} \\

\bottomrule
\end{tabular}
\end{table}

\subsection{Semantic-to-Mel Architecture Ablation}
We conduct a subjective preference evaluation between U-DiT and Zipformer as backbone architectures for the S2M module, using a held-out test subset. Listeners compare generated audio pairs along three dimensions: speech quality, speaker similarity, and prosodic naturalness.

\begin{figure}[H]
    \centering
    \includegraphics[width=0.9\textwidth]{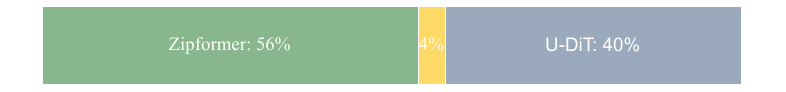}
    \caption{Subjective preference comparison of U-DiT and Zipformer backbone architectures in the S2M module.}
    \label{fig:abtest}
\end{figure}

As shown in Figure~\ref{fig:abtest}, the system employing the Zipformer-based backbone is preferred in 56\% of trials, compared to 40\% for the U-DiT-based variant, with 4\% of responses indicating no clear preference. This consistent preference suggests that the Zipformer-based architecture generates more natural and speaker-consistent speech, likely due to its efficient modeling of long-range dependencies and stable training dynamics. The results highlight that, despite its simpler design, the Zipformer-based flow-matching module achieves stronger perceptual performance than the more complex U-DiT counterpart in zero-shot multilingual TTS, offering a favorable trade-off between model capacity and synthesis quality.

\subsection{Inference Performance Comparison}
\begin{table}[H]
\centering
\caption{Inference speed comparison of IndexTTS 2 series models.}
\label{tab:rtf} 
\begin{tabular}{@{}lcc@{}}
\toprule
\multirow{2}{*}{\textbf{Model}} & \multicolumn{2}{c} {\textbf{RTF$\downarrow$}} \\
\cmidrule(lr){2-3} & \textbf{T2S} & \textbf{S2M}  \\
\midrule
\multirow{1}{*}{\textbf{IndexTTS 2}}
& 0.232 & 0.078  \\
\midrule
\multirow{1}{*}{\textbf{IndexTTS 2.5 (U-DiT)}}
& 0.119 & 0.081  \\
\midrule
\multirow{1}{*}{\textbf{IndexTTS 2.5 (Zipformer)}}
& 0.119 & 0.017  \\
\midrule
\end{tabular}
\end{table}
Table~\ref{tab:rtf} summarizes the inference efficiency of the IndexTTS 2 series under identical hardware conditions (NVIDIA A10 GPU and Intel Xeon Platinum 8350C CPU). IndexTTS 2.5 achieves substantially lower RTF in both the T2S and S2M modules compared to IndexTTS 2. The RTF reduction in T2S—from 0.232 to 0.119—stems from frame-rate compression in the neural codec, which shortens the semantic token sequence while preserving linguistic fidelity. In parallel, architectural improvements in the S2M module, particularly the adoption of the Zipformer backbone, reduce its RTF to 0.017. These optimizations enable faster inference without degrading synthesis quality.

\section{Conclusion}

IndexTTS 2.5 enhances zero-shot multilingual emotional text-to-speech synthesis through four core improvements: semantic codec compression, an optimized S2M architecture based on Zipformer, cross-lingual modeling strategies, and reinforcement learning–based T2S optimization. Experimental results show that these enhancements jointly enable high-quality synthesis across Chinese, English, Japanese, and Spanish, preserving both speaker identity and emotional prosody in unseen languages. While achieving a 2.28× reduction in real-time factor, the system maintains low WER and high SS. These results demonstrate that IndexTTS 2.5 achieves an effective balance between inference efficiency and synthesis quality, making it suitable for real-world deployment of expressive multilingual speech synthesis.

\section{Contributors}
The contributors are listed in order of contribution:

Yunpei Li, Xun Zhou, Jinchao Wang, Lu Wang, Yong Wu, Siyi Zhou, Yiquan Zhou, Yining Wang, Yaogen Yang, Zhetao Hu, Shiyao Duan, Jiacheng Xu, Jingchen Shu, Bin Xia
{
\small
\nocite{*}
\bibliographystyle{IEEEtran}
\bibliography{ref.bib}
% \end{document}

}

\end{document}